\documentclass[aps,prl,reprint]{revtex4-1}

\usepackage{graphicx}
\usepackage{dcolumn}
\usepackage{bm}
\usepackage{amsmath}
\usepackage{amsfonts}
\usepackage{amssymb}
\usepackage{braket}

\let\oldsqrt\sqrt
\def\sqrt{\mathpalette\DHLhksqrt}
\def\DHLhksqrt#1#2{%
\setbox0=\hbox{$#1\oldsqrt{#2\,}$}\dimen0=\ht0
\advance\dimen0-0.2\ht0
\setbox2=\hbox{\vrule height\ht0 depth -\dimen0}%
{\box0\lower0.4pt\box2}}

\begin{document}

\preprint{LR15153}

\title{Simulating confined particles with a flat density profile}

\author{Airidas Korolkovas}
\email{korolkovas@ill.fr}
\affiliation{Institut Laue-Langevin, 71 rue des Martyrs, 38000 Grenoble, France}
\affiliation{Universit\'{e} Grenoble Alpes, Liphy, 140 Rue de la Physique, 38402 Saint-Martin-d'H\`{e}res, France}

%
%

\date{\today}

\begin{abstract}
Particle simulations confined by sharp walls usually develop an oscillatory density profile. For some applications, most notably soft matter liquids, this behavior is often unrealistic and one expects a monotonic density climb instead. To reconcile simulations with experiments, we propose mirror-and-shift boundary conditions where each interface is mapped to a distant part of itself. The main result is that the particle density increases almost monotonically from zero to bulk, over a short distance of about one particle diameter. The method is applied to simulate a polymer brush in explicit solvent, grafted on a flat silicon substrate. The simulated density profile agrees favorably with neutron reflectometry measurements and self-consistent field theory results.
\end{abstract}

\maketitle

\section{Introduction}

Liquids at interfaces are ubiquitous: functionalized surfaces, lipid membranes, microphase separation, pores and cavities, microfluidic devices, and just about any other place where a liquid comes into contact with a solid, a gas, or another incompatible liquid. The fluid properties close to the wall are often markedly different from the bulk and it is interesting to study them using computer simulations. In particle-based models one must usually apply some confinement force at the interface and in doing so break the homogeneity of space. This invariably results in an oscillatory density profile with large deviations from the bulk level. For some systems, such as hard liquids at very crisp walls, the oscillations are realistic and match experimental data\cite{cheng2001molecular, mo2005ordering}. In other situations, most notably coarse-grained softer fluids, the experimental density profile is usually flat\cite{maccarini2007water}, and so the simulated oscillations should be seen as an unwanted computational artifact. In this letter we show how to suppress such modeling bias and provide a better agreement between experiment and simulation. The main idea is to avoid border discontinuities altogether by mapping each interface back onto a distant part of itself, via mirror-and-shift boundary conditions.

\section{Model system}
Our prototype system will be a liquid soft matter sample made up of large molecules with a characteristic size in the range of $\lambda \in (1;\, 100)\, \text{nanometers}$. This length scale describes the dominant features for colloids, dendrimers, polymers, surfactants, etc. The liquid is placed on a flat surface, such as a silicon wafer, with roughness $\sigma \in (0.1;\, 1)\, \text{nm} \ll \lambda$, much smaller than the size of the soft particles. In this scenario the chemical details of the interface are not important and cannot possibly be resolved by the coarse calculations. Therefore, the example is equally applicable to liquid/air, liquid/solid, or liquid/liquid interfaces.

\begin{figure}[bht]
\centering
    \includegraphics[width=0.9\linewidth]{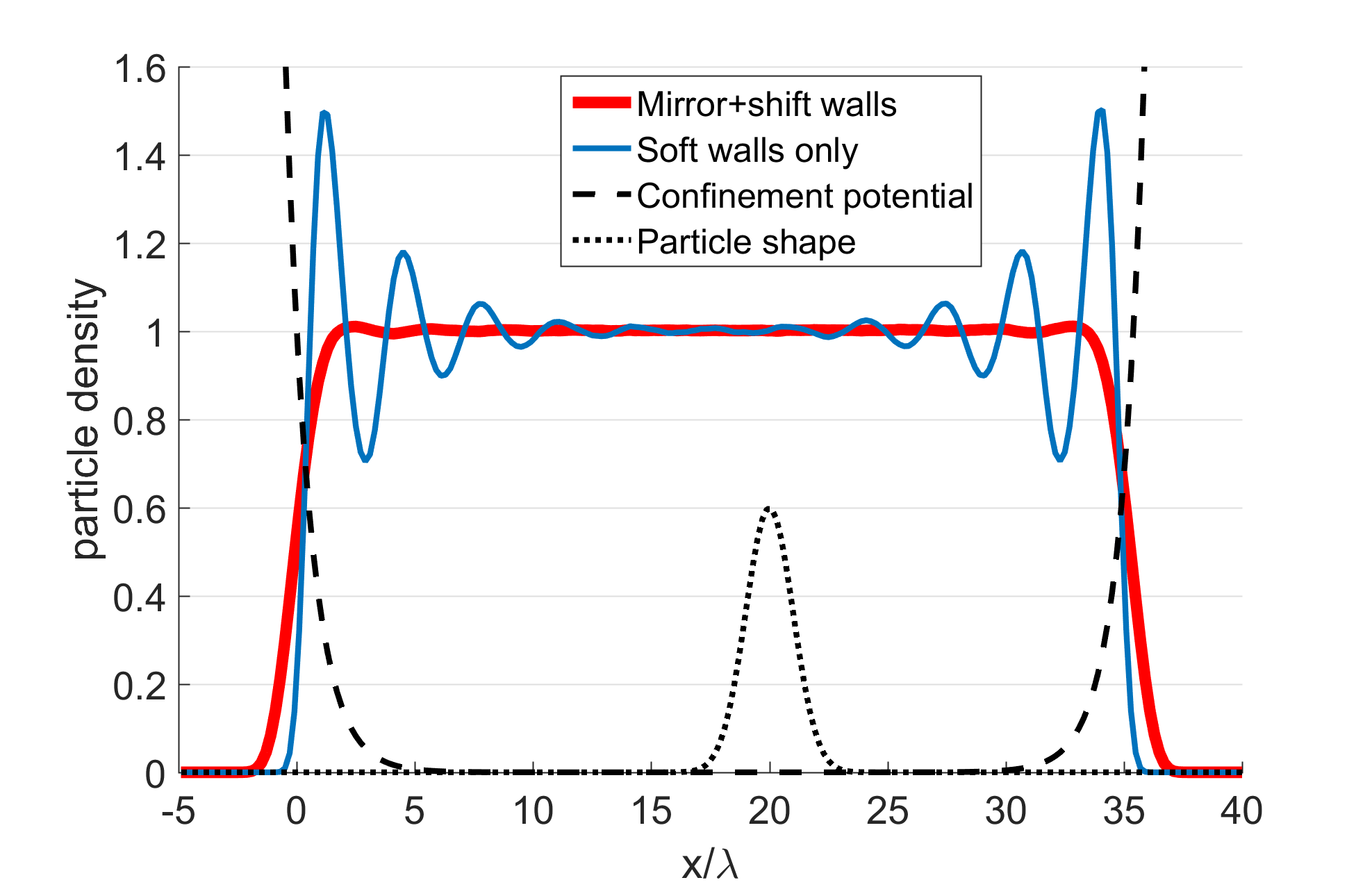}
    \caption{Density of soft repulsive particles, closely packed between two walls is shown by the thin blue line. The oscillations can be suppressed by using mirror-and-shift boundary conditions, without compromising the sharpness of the interface, as shown by the thick red line. The dashed line shows the confinement potential $U(x)$ and the dotted line shows the particle shape $\Phi(\mathbf{r})$, on an arbitrary $y$-scale. }\label{density}
\end{figure}

A number $N$ of particles is closely packed in a box of volume
\begin{equation}\label{denseeq}
V = L^3 = \frac{4\pi}{3} N\lambda^3,
\end{equation}
and in the case of polymers this corresponds to the cross-over density between dilute and semi-dilute: $\rho=\rho^*$. It is clear from experiment\cite{thomas1996neutron} that the particle concentration increases monotonically from zero outside the box, to $N/V$ inside the box, and in a simulation we should ideally be able to resolve the climb within one particle diameter $2\lambda$, as shown with the thick red line in Figure~\ref{density}. 

Surprisingly, it is quite difficult to achieve this behavior with a particle-based model, and most attempts will lead to an oscillatory density profile shown with a thin blue curve in Figure~\ref{density}. As a concrete example, let the particles repel each other by a Gaussian potential
\begin{equation}\label{gaussian}
\Phi (\mathbf{r}) = \epsilon \exp \left(-\frac{\mathbf{r}^2}{2\lambda^2}\right),
\end{equation}
where $\epsilon \gg k_B T$ is the repulsive strength chosen so that the particles cannot cross each other. The most common way to model a confining surface is to impose an external field $U(x)$. It will not be possible to resolve small surface features $\sigma$ using particles of large size $\lambda$, and so the sharpest confinement could have its roughness equal to the particle size, $ |\partial U/\partial x| \approx U/\lambda$, which leads to
\begin{equation}\label{ufield}
U(x) = u \left( e^{-x/\lambda} + e^{-(L-x)/\lambda}\right)
\end{equation}
with the confinement strength parameter $u$ adjusted so that the density in the middle of the box equals $N/V$, to fulfill Equation~\eqref{denseeq}. The resulting density profile is oscillating as shown with the thin blue curve in Figure~\ref{density}. Similar oscillations occur with various other shapes of the confinement $U(x)$ and the excluded volume force $\Phi(\mathbf{r})$. The oscillations do not decrease in larger $N\rightarrow \infty$ systems. They also persist for more complex fluids containing rods, chains, or polydisperse mixtures of various shapes. In general, the density oscillations are always present on the scale of the smallest particle size $\lambda$, provided that the surface roughness is on the same scale or smaller: $\sigma \lesssim \lambda$.

\begin{figure}[bht]
\centering
    \includegraphics[width=0.9\linewidth]{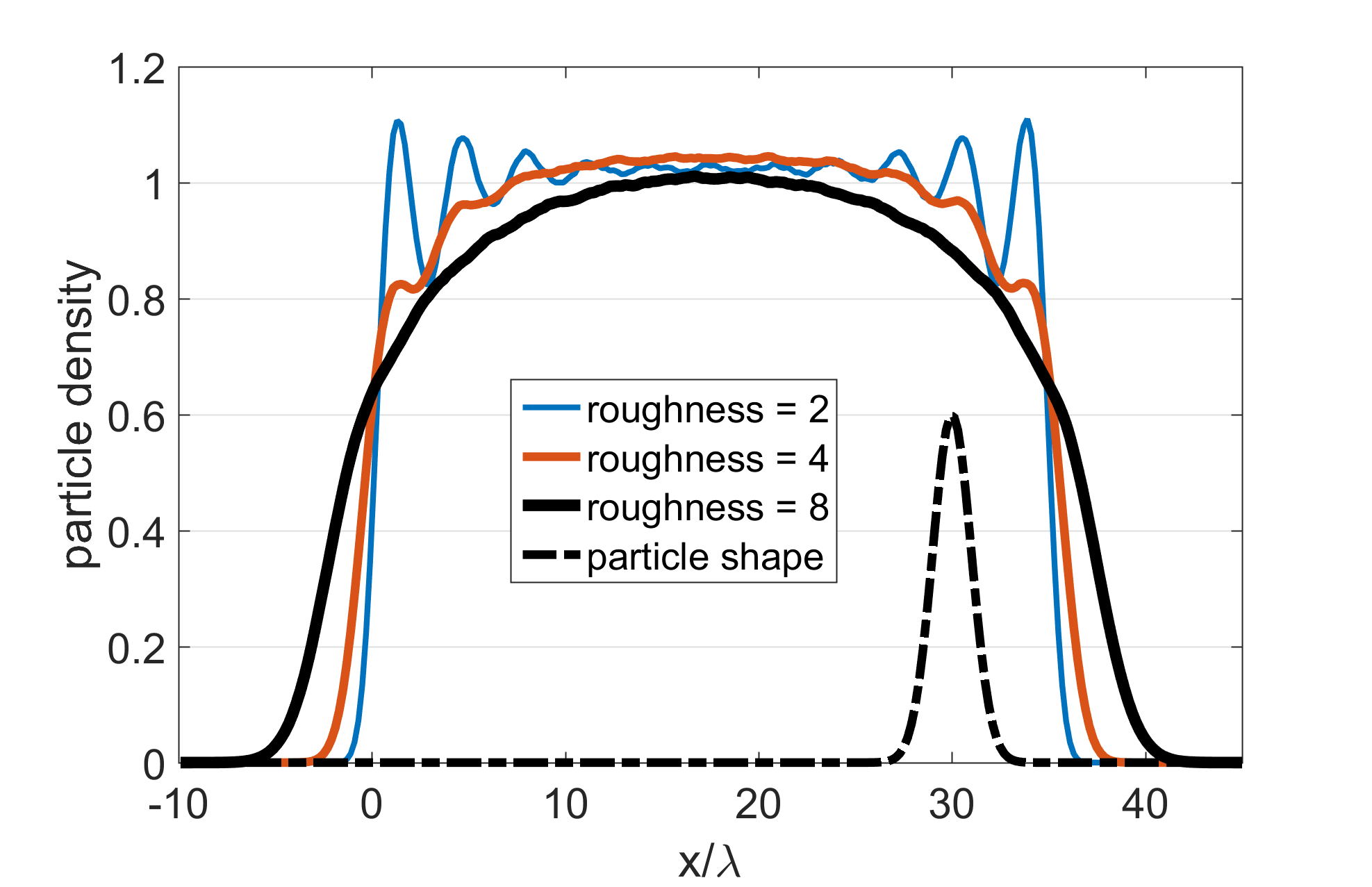}
    \caption{Density profile dependence on surface roughness}\label{roughness}
\end{figure}

The most obvious way to avoid the oscillations is to assume a large surface roughness $\sigma\gtrsim 6\lambda$, as demonstrated in Figure~\ref{roughness}. Unfortunately, this assumption is not valid in many important applications, most notably soft matter at flat walls: $\lambda \gg \sigma$. This issue has been recognized in the literature before and several studies have reported alternative methods of confinement. The main idea\cite{pivkin2006controlling, kotsalis2007control, issa2014algorithm} was to invert the density oscillations by applying a counter-oscillating confinement force. While such an approach does deliver a flat density profile, there remain some shortcomings: the confinement force has to extend many $\lambda$'s deep into the bulk of the box; the detailed shape of the force must be recalculated for every different fluid chemistry; the algorithms are somewhat complicated and unnatural.

\begin{figure}[bht]
\centering
    \includegraphics[width=0.4\linewidth]{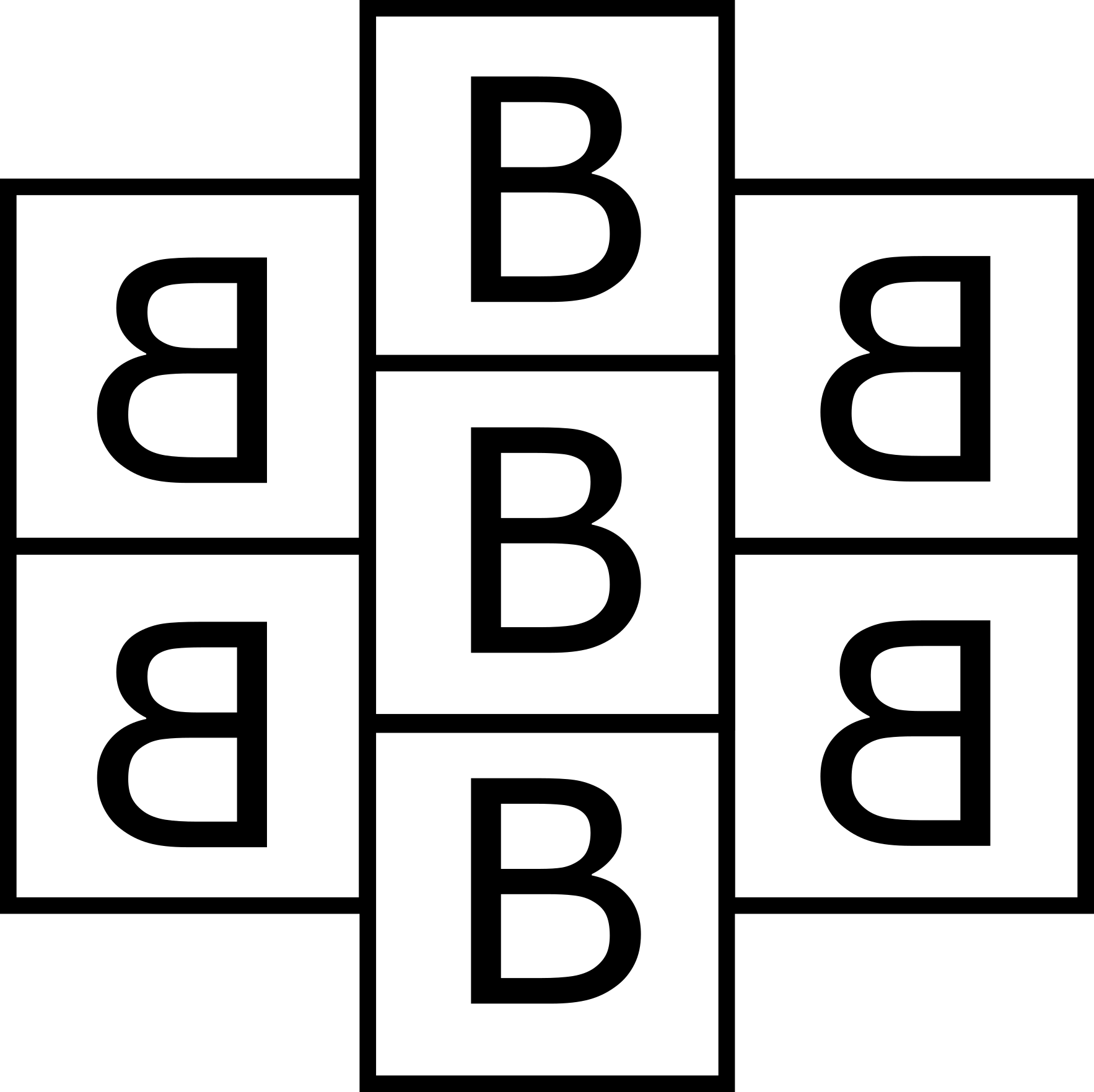}
    \caption{Mirror-and-shift boundary condition}\label{mirror}
\end{figure}

In this work we explore a new strategy aimed to alleviate some of these problems. Our main contribution is to notice that an exact counter-oscillating confinement force is already available to us, without requiring any further models or calculations. We simply map each interface back onto itself by a mirror transformation as shown in Figure~\ref{mirror}. It is also crucial to shift the mirror images along the walls to avoid the particles interacting with themselves, as this would create unnatural correlations which behave differently from the bulk. This mapping equalizes most of the osmotic pressure across the boundary, and the only force left to drive the particles out of the box is just random diffusion. Fortunately, the energy of a random walk is only one thermal unit $k_B T\ll \epsilon$, which is small and can be repelled back to the main box by the same external field $U(x)$ from Eq.~\eqref{ufield}, but with a weak amplitude $u\approx 1\, k_B T$. The system perceives this repulsion as a small perturbation to an otherwise perfectly balanced system, and as a result, the density oscillations are almost entirely suppressed, whereas the width of the boundary remains just a few $\lambda$'s which is about as sharp as one can hope for.

\section{Method}
To dig straight to the root of the problem, we propose a very simple model of a fluid confined between two walls. We will work in two dimensions and so the box width is set to $L = \sqrt{4\pi \lambda^2 N}$. The position of each particle $\mathbf{R}_n$ evolves according to the Brownian equation of motion
\begin{equation}\label{eqmotion}
\zeta \frac{\partial \mathbf{R}_n}{\partial t} = \sqrt{2\zeta k_B T } \mathbf{W}_n(t) - \mathbf{F}_n
\end{equation}
where $\braket{\mathbf{W}_n^{\alpha}(t)\mathbf{W}_m^{\beta}(t')} = \delta^{\alpha \beta}\delta_{nm}\delta(t-t')$ is the Wiener process and
\begin{equation}
\mathbf{F}_n = -\nabla \left(U(\mathbf{r}) + \sum_{m=1}^N \Phi(\mathbf{r}-\mathbf{R}_m)  \right) \Big|_{\mathbf{r} = \mathbf{R}_n}
\end{equation}
is the confinement plus the interparticle forces. Care has been taken to resolve the equation of motion very accurately to ensure that the oscillations at the wall are an actual property of the model and not some discretisation error. Integration over a short time $\Delta t$ using the mid-point rule gives
\begin{multline}
\mathbf{R}_n^{\texttt{(i)}} (t+\Delta t) = \mathbf{R}_n (t) + \mathcal{R}_n \sqrt{\frac{2k_B T\Delta t}{\zeta}} +\\
+ \frac{\Delta t}{\zeta} \left(\frac{\mathbf{F}_n (t) + \mathbf{F}_n^{\texttt{(i)}}(t+\Delta t)}{2}\right).
\end{multline}
The random vector $\mathcal{R}_n$ has a fixed radius of $\sqrt{2}$ and a random orientation $\theta \in (0; 2\pi)$. Since the force at the end of the time step $\mathbf{F}_n(t+\Delta t)$ is not known, we initially guess $\mathbf{F}_n^{\texttt{(1)}}(t+\Delta t) = \mathbf{F}_n(t)$, and then evaluate all the forces again using the $\mathbf{R}_n^{\texttt{(1)}}$ configuration to obtain a better guess $\mathbf{F}_n^{\texttt{(2)}}(t+\Delta t)$. This process is repeated until the residual $\texttt{max} |\mathbf{R}_n^{\texttt{(i)}}-\mathbf{R}_n^{\texttt{(i-1)}}|<0.01 \lambda$ is only a tiny fraction of the particle size. The dimensionless time step $\lambda^2 \Delta t/\zeta \approx 1/\epsilon$ is chosen so that the above convergence criterion is met in $\texttt{i} \approx 3$ iterations on average. 

The first run was simulated on a system with $N=100$ particles, $\epsilon=100 k_B T$ interparticle barrier and $u=6.5 k_B T$ confinement strength. Periodic boundary condition was applied on the $y$-axis:
\begin{equation}\label{periodic}
y_{nm} \rightarrow y_{nm} - L\, \texttt{round}\left(y_{nm}/L\right),
\end{equation}
and no such condition on the confined $x$-axis. The resulting long-time average density profile is shown with a blue oscillating line in Figure~\ref{density}.

In the second run we have simulated the same system but included additional interparticle pairs. For every particle $\mathbf{R}_m$ we construct its mirror-and-shifted counterparts on the left and right sides of the box:
\begin{equation}
\mathbf{R}_{m}^{\text{(L)}} = 
\begin{pmatrix}
 -x_{m}\\ y_{m} + L/2
\end{pmatrix} \quad \text{and} \quad
\mathbf{R}_{m}^{\text{(R)}} = 
\begin{pmatrix}
 2L-x_{m}\\ y_{m} + L/2
\end{pmatrix}
\end{equation}
and then every original particle $\mathbf{R}_n$ feels three sets of interactions: $\mathbf{R}_{nm}^{(L,0,R)} = \mathbf{R}_n - \mathbf{R}_m^{(L,0,R)}$. The usual periodic boundary condition on the $y$-axis is afterward applied to all the pairs, using Eq.~\eqref{periodic}. In short, every particle $\mathbf{R}_n$ interacts with all the other particles $\mathbf{R}_m$ as well as their both mirror-shifted images as defined above.

An external field from Equation~\eqref{ufield} is still required to confine random diffusion, but now a much weaker barrier $u = 1\, k_B T$ was sufficient to keep the density in the center of the box at the nominal level $N/V$ as designated by Eq.~\eqref{denseeq}. The density profile thus obtained is shown by the thick red line in Figure~\ref{density}. Admittedly, there still remains a slight, about 1\% high, overshoot after the boundary, but it is a considerable reduction from 50\% overshoot which is obtained without the use of mirrors.


\section{Application: polymer brush}
\begin{figure}[bht]
\centering
    \includegraphics[width=0.8\linewidth]{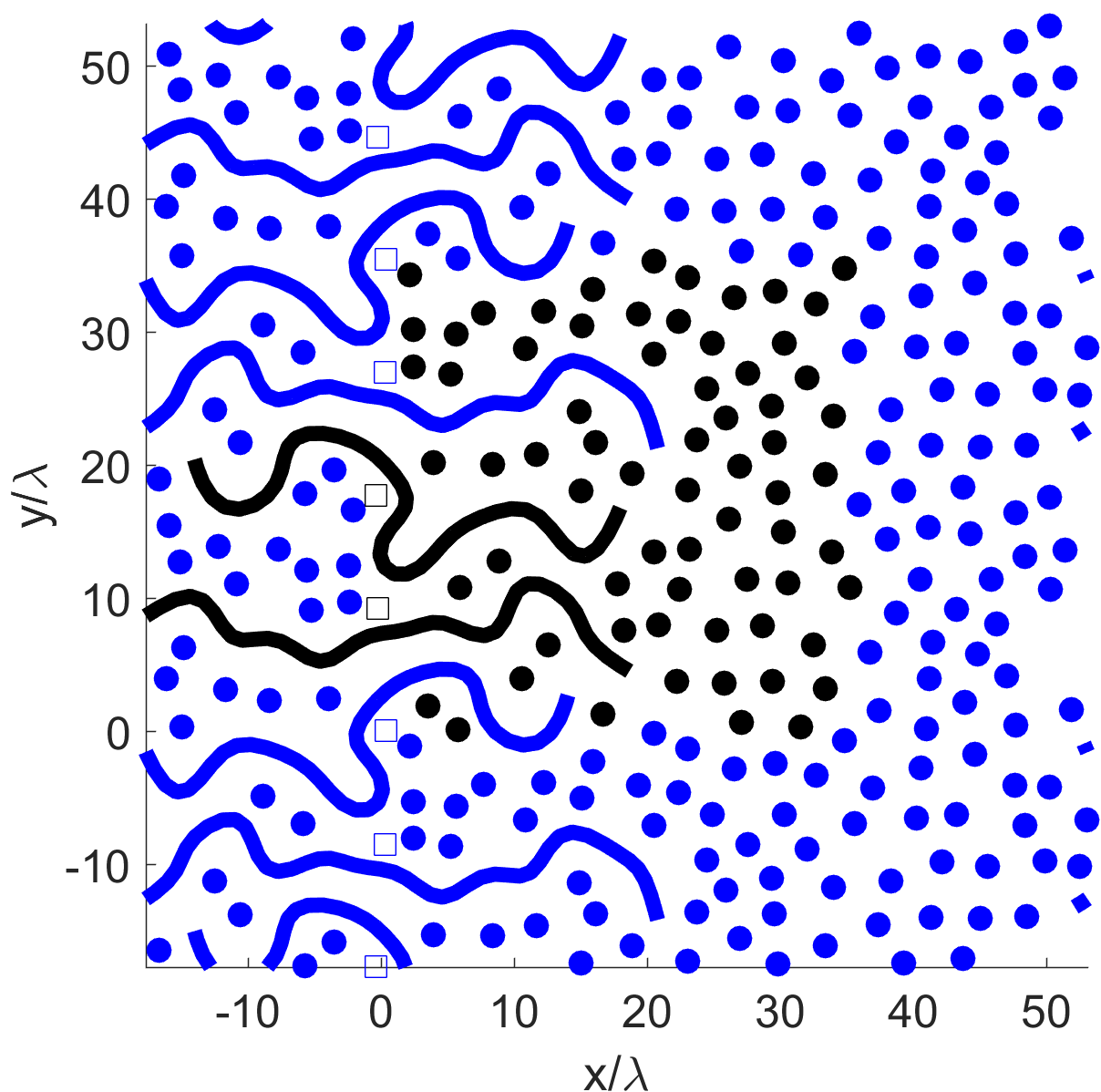}
    \caption{Polymer brush in a good solvent. The main box is the black region $\mathbf{r}\in (0;\, 35.4)^2$. Grafting points are shown by hollow squares.}\label{config2}
\end{figure}

To demonstrate the usefulness of our new method, we present one real-world application whose simulation necessitates confinement. We will study a polymer brush grafted on a flat silicon wafer and immersed in a good solvent, as shown in Figure~\ref{config2}. It is essentially the same simulation as in the previous section, with the solvent molecules now coarse-grained into blobs of size $\lambda$, and $N=16$ of the blobs are connected into a chain by linear springs of length $b=\lambda$. We have added $C=2$ of such chains on the $x=0$ interface, corresponding to a grafting density $\sigma = 2C\lambda/L = 0.11$. The chains are overlapping and are in the stretched brush regime.\cite{de1980conformations}

A detailed description of the pseudo-continuous chain model is given in our previous work\cite{korolkovas2016}, and here we will only mention the essential equations. To suppress the likelihood of the chains crossing each other, the polymer is described by a continuous curve with $N$ degrees of freedom:
\begin{equation}
\mathbf{R}(s) = \mathbf{a}_0 + 2\sum_{n=1}^{N-1} \mathbf{a}_n \cos ( \pi s n ), \quad s\in (0;\, 1)
\end{equation}
which is sampled at discrete points $j=1,\, 2,\, \ldots,\, J\geq N$,
\begin{equation}\label{Rj}
\mathbf{R}_j = \mathbf{a}_0 + 2\sum_{n=1}^{N-1} \mathbf{a}_n \cos \left[ \pi \left(\frac{2j-1}{2J}\right)n\right],
\end{equation}
and we have used $J=3N+1=49$, but the precise choice is not important as long as $J\gg N$. The motion of every $j$-particle follows the same Equation~\eqref{eqmotion} as the solvent particles, with a few minor alterations. First, the intermolecular potential stemming from a $j = (Js_0 + 1/2)$ particle is
\begin{align}
\Phi_j (\mathbf{r}-\mathbf{R}_j) &= \int_{s_0 - 1/(2J)}^{s_0 + 1/(2J)} ds\, N \Phi (\mathbf{r}-\mathbf{R}(s))\\
& \approx \left(\frac{N}{J}\right) \Phi(\mathbf{r}-\mathbf{R}_j)
\end{align}
reduced by $N/J$. The friction coefficient is also decreased: $\zeta_j = (N/J) \zeta$, but so is the coupling to the potential field: $\mathbf{F}_j = -(N/J)\nabla (U + \sum \Phi)$. A linear spring force
\begin{equation}
\mathbf{F}_{\text{spring}} = \left(\frac{3k_B T}{Nb^2} \right) \frac{\partial^2 \mathbf{R}}{\partial s^2}
\end{equation}
is included by first going to the Rouse representation
\begin{align}
\mathbf{a}_n &= \int_0^1 ds\, \mathbf{R}(s) \cos (\pi sn)\\
&\approx \frac{1}{J}\sum_{j=1}^J \mathbf{R}_j \cos \left[ \pi \left(\frac{2j-1}{2J}\right)n\right],
\end{align}
and then applying the Backwards Euler scheme:
\begin{equation}
\mathbf{a}_n \rightarrow \frac{\mathbf{a}_n}{1+3\pi^2 \left(\frac{k_B T \Delta t}{\zeta b^2}\right) \left(\frac{n}{N}\right)^2}.
\end{equation}
At this point, the configuration in real space $\mathbf{R}_j $ is recovered by Eq.~\eqref{Rj}, and the next iteration can begin.

The chains are uniformly grafted at $x=0$ plane, where ``uniformly'' means at equidistant points, plus a random number of variance $\lambda$ in all directions. The grafting model should at the very least: 1) keep the interface as homogeneous as possible, to avoid density oscillations, and 2) ensure that the solvent particles do not sneak in between the grafted chain and the wall. One way to satisfy these requirements is to anchor the central point $s_0=1/2 \rightarrow j_0=(J+1)/2$ to its designated grafting location using a soft attractive potential
\begin{equation}
U_{\text{graft}}(\mathbf{r}) = -k_B T \cosh (\mathbf{r}/\lambda).
\end{equation}
Next, the first half of the chain $s<1/2$ is assigned to the main box, and feels the wall $U(x)$ from Eq.~\eqref{ufield}, just like all the solvent particles. The second half of the chain $s>1/2$ is assigned to the mirror box where it feels the reflected wall $U(-x)$. A snapshot of the resulting conformation is plotted in Figure~\ref{config2}, showing that each chain plays a tug of war between the main box and its shifted mirror, contributing two bristles of length $N/2$ to the brush.

\begin{figure}[bht]
\centering
    \includegraphics[width=0.9\linewidth]{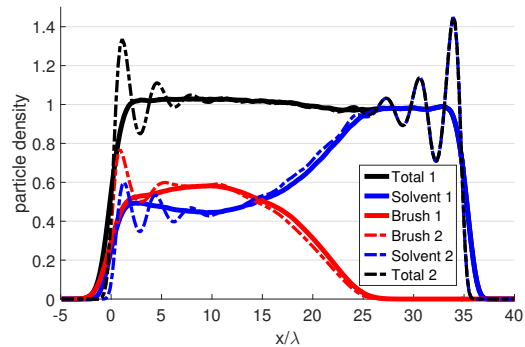}
    \caption{Brush density profile. Solid lines result from mirror-and-shift walls, whereas dashed oscillating lines originate from external confinement alone.}\label{scft}
\end{figure}

The density profile of the brush and the solvent is shown in Figure~\ref{scft}. Notice that the polymer is slightly denser than the chemically identical solvent, which is to be expected because of the spring attraction. Our result is quite similar to the self-consistent field theory calculations\cite{zhulina1991theory} as well as experimental data of PDMS\cite{marzolin2001neutron} and polystyrene\cite{devaux2005low} brushes swollen in toluene and measured using neutron reflectometry. Most importantly, the experiments decidedly exclude density oscillations of 50\% which would be present in a simulation without mirror walls (dashed lines), and as seen in other simulation studies\cite{kreer2016polymer}.



\section{Conclusion}
In general, particles under confinement will develop density oscillations if the surface roughness $\sigma$ is not substantially greater than the smallest particle size $\lambda$. This has been a common problem for coarse-grained simulations where the confining surface can be atomically crisp, while the liquid ``particles'' are large clusters of many atoms, grouped together for computational convenience. In most experiments the density profile is monotonic, whereas the majority of simulations report an oscillating density.

In this study we have succeeded to suppress the oscillations considerably by mirroring the interface back onto a distant part of itself, so that the system self-equilibrates and remains quasi-homogeneous at the boundary. The only imperfection is that we must still add a weak external confinement to counteract thermal diffusion, but this is a small perturbation and the remaining density oscillations are tiny. 

The mirror-and-shift boundary conditions do not require any input parameters and are scale-independent. Therefore, they could in principle be applied to any other situation, as a replacement of the usual periodic boundary conditions. One could also use the method to just alter the topological connectivity of the simulation box, without adding the confining field.

\section{Acknowledgments}
The author thanks Jean-Louis Barrat and Philipp Gutfreund for critical comments during the preparation of this manuscript.



\bibliography{manuscript}

\end{document}